\documentclass[iop]{emulateapj}

\shorttitle{ROTATIONAL INSTABILITY IN OUTER DISKS}
\shortauthors{Ono et al.}

\begin{document}

%% LaTeX will automatically break titles if they run longer than
%% one line. However, you may use \\ to force a line break if
%% you desire.

\title{ROTATIONAL INSTABILITY IN THE OUTER REGION OF PROTOPLANETARY DISKS}

%% Use \author, \affil, and the \and command to format
%% author and affiliation information.
%% Note that \email has replaced the old \authoremail command
%% from AASTeX v4.0. You can use \email to mark an email address
%% anywhere in the paper, not just in the front matter.
%% As in the title, use \\ to force line breaks.

\author{Tomohiro Ono, }
\affil{Department of Astronomy, Graduate School of Science, Kyoto University, Sakyo-ku, Kyoto 606-8502, Japan}
\email{ono.t@kusastro.kyoto-u.ac.jp}

\author{Hideko Nomura, AND Taku Takeuchi}
\affil{Department of Earth and Planetary Sciences, Tokyo Institute of Technology, Ookayama, Meguro-ku, Tokyo 152-8551, Japan}
%% Notice that each of these authors has alternate affiliations, which
%% are identified by the \altaffilmark after each name.  Specify alternate
%% affiliation information with \altaffiltext, with one command per each
%% affiliation.

%% Mark off your abstract in the ``abstract'' environment. In the manuscript
%% style, abstract will output a Received/Accepted line after the
%% title and affiliation information. No date will appear since the author
%% does not have this information. The dates will be filled in by the
%% editorial office after submission.

\begin{abstract}
We analytically calculate the marginally stable surface density profile for the rotational instability of protoplanetary disks. The derived profile can be utilized for considering the region in a rotating disk where radial pressure gradient force is comparable to the gravitational force, such as an inner edge, steep gaps or bumps, and an outer region of the disk. In this paper, we particularly focus on the rotational instability in the outer region of disks. We find that a protoplanetary disk with a surface density profile of similarity solution becomes rotationally unstable at a certain radius, depending on its temperature profile and a mass of the central star. If the temperature is relatively low and the mass of the central star is high, disks have rotationally stable similarity profiles. Otherwise, deviation from the similarity profiles of surface density could be observable, using facilities with high sensitivity, such as ALMA.
\end{abstract}

%% Keywords should appear after the \end{abstract} command. The uncommented
%% example has been keyed in ApJ style. See the instructions to authors
%% for the journal to which you are submitting your paper to determine
%% what keyword punctuation is appropriate.


%% From the front matter, we move on to the body of the paper.
%% In the first two sections, notice the use of the natbib \citep
%% and \citet commands to identify citations.  The citations are
%% tied to the reference list via symbolic KEYs. The KEY corresponds
%% to the KEY in the \bibitem in the reference list below. We have
%% chosen the first three characters of the first author's name plus
%% the last two numeral of the year of publication as our KEY for
%% each reference.

%% Authors who wish to have the most important objects in their paper
%% linked in the electronic edition to a data center may do so by tagging
%% their objects with \objectname{} or \object{}.  Each macro takes the
%% object name as its required argument. The optional, square-bracket 
%% argument should be used in cases where the data center identification
%% differs from what is to be printed in the paper.  The text appearing 
%% in curly braces is what will appear in print in the published paper. 
%% If the object name is recognized by the data centers, it will be linked
%% in the electronic edition to the object data available at the data centers  
%%
%% Note that for sources with brackets in their names, e.g. [WEG2004] 14h-090,
%% the brackets must be escaped with backslashes when used in the first
%% square-bracket argument, for instance, \object[\[WEG2004\] 14h-090]{90}).
%%  Otherwise, LaTeX will issue an error. 

\section{Introduction}
Protoplanetary disks evolve via turbulent viscosity, and their evolution has been well represented by the model of \citeauthor*{Lyn74}(1974, henceforth LBP74). 
In the LBP74 model, the rotation profile is assumed to be time-independent or simply Keplerian. 
This assumption is justified if the radial pressure gradient is much smaller than the central star gravity. 
The temperature in the disks is generally low enough that the radial pressure gradient can be neglected in most parts of the disk. 
However, even in disks with low temperature, the radial pressure gradient may not be neglected in which the surface density dramatically varies in the radial direction. 
Such regions appear, for example, at the gap edge formed by the gravity of planets, and at the boundary between the active and inactive regions for magneto-rotational instability. 
For such regions, we need to modify the LBP74 model to take into account the change in the rotation profile. 
Interestingly, the similarity solution of the LBP74 model has an exponential cut off at the outermost region (LBP74; \citealp{Har98}). 
In the outer edge of the disk, the density decreases exponentially with a radial scale length, R0. 
Because the disk scale height, H, generally increases with the radius, the scale height becomes larger than the radial scale length at a certain point, resulting in a violation of the assumption of negligible pressure gradient. 
Hence, even if there is no external force modifying the disk structure, such as photoevaporation or planets, the self-similar solution of the LBP74 model violates self-consistency at the outer edge of the disk.

The recent development of high-sensitivity (sub)millimeter interferometers makes it possible to observe surface density profiles of the outer regions of protoplanetary disks. 
The observations have suggested that the profiles are well fitted by a similarity solution of the disks (e.g., \citealp{Hug08}; \citealp{And09}, \citeyear{And10}; \citealp{Aki13}), which has a sharp density profile at the outer edge. 
Observations using facilities with high sensitivity, such as ALMA, however, are expected to reveal the lower density area in the outer region in which the radial pressure gradient force is non-negligible. 
Because of this, it is important that we investigate the outer region of the disks without the assumption of Keplerian disks.

When the gas pressure gradient force becomes non-negligible compared with the gravitational force in the equation of motion in the radial direction, an assumption of Keplerian rotation becomes inadequate.
When we inappropriately adopt the assumption to a disk with a steep surface density profile and a large radial pressure gradient force, we often see rotational instability.
Rotational instability is one of the hydrodynamical instabilities in axisymmetric differentially rotating disks \citep{Cha61}. 
The Rayleigh's criterion is the discriminant for rotational stability; for an inviscid disk to be rotationally stable, the specific angular momentum ($j$) must monotonically increase with cylindrical distance from the axis of rotation ($R$) in a flow:
\begin{equation}
\frac{\partial j^2}{\partial R}> 0. \label{1}
\end{equation}
Otherwise, the disk becomes rotationally unstable and the gas
radially migrates with conserving specific angular momentum.
In a viscid disk, viscosity limits the onset of rotational instability.
Also, in the Rayleigh's criterion, 
the radial entropy gradient is not taken into account. 
We need to use the Solberg-Hoiland criterion \citep{End78} for
a disk with a radial entropy gradient, with which the rotational
instability is easier to set in standard disks \citep{Sha73}. 
In this paper, however, we simply adopt the Rayleigh's criterion as the discriminant for rotational stability because we treat disks with low viscosity and the Rayleigh's criterion is a more severe discriminant for rotational instability than the Solberg-Hoiland criterion.

Since the sharp edge could lead to rotational instability,  in this paper, we investigate the condition under which the disk becomes rotationally unstable.
We also analytically calculate the marginally stable surface density profile for rotational instability, which indicates that the profile becomes shallower than that of the similarity profile.
The rotation velocity will be less than the Keplerian velocity in the region where the disk is rotationally unstable. 
If the deviation from the Keplerian velocity is observable, it will be evidence that the radial pressure gradient force is sufficiently strong in the region.

We analytically examine the marginally stable disks for rotational instability in Section 2. 
In Section 3, we apply the result of Section 2 for protoplanetary disks. 
We discuss the possibility of observations of non-similarity profiles and the validity of approximations used in this work in Section 4, and we summarize our conclusions in Section 5.
Although many discussions here are applicable to the outer region of
accretion disks in general, we focus on protoplanetary disks in this paper.

\section{Marginally Stable Condition for Rotational Instability}
In this section, we analytically calculate the marginally stable surface
density profile for rotational instability.
Then we connect the marginally stable profile to the similarity profile of viscous disks.

\subsection{Rayleigh's Criterion for Protoplanetary Disks}
We consider an axisymmetric rotating disk with hydrostatic equilibrium in the vertical direction. 
The equation for radial force balance in cylindrical coordinates is 
\begin{eqnarray}
\frac{j^2}{R^3} &=& \frac{GM}{R^2}+\frac{1}{\rho} \frac{\partial P}{\partial R}, \nonumber \\
 &=& \frac{GM}{R^2}+\frac{H}{\Sigma} \frac{\partial (c^2_s \Sigma /H)}{\partial R}, \nonumber \\
 &=& \frac{GM}{R^2}+\frac{1}{\Sigma} \frac{\partial (c^2_s \Sigma)}{\partial R}- \frac{c^2_s}{H} \frac{\partial H}{\partial R}, \label{2} 
\end{eqnarray}
where $G$ is the gravitational constant, $M$ the mass of the central
star, $P$ the pressure, $\rho$ the density at midplane, $\Sigma$ the surface
density, $H\equiv c_s/\Omega_K$ the scale height, $c_s$ the sound speed
of the gas, and $\Omega_K\equiv (GM/R^3)^{1/2}$ the Keplerian angular velocity.
We adopt the isothermal equation of state, $P=c_s^2\rho$. 
We assume that the disk is geometrically thin $H/R \ll 1$, and that the vertical structure is isothermal, i.e., \mbox{$\Sigma=\sqrt{2 \pi} \rho H$}.

For generalization, we define normalized non-dimensional
parameters for the disk radius, $r$, and the surface density, $\sigma$,
as follows, 
\begin{equation}
R = R_0 r \ \ {\rm and} \ \  \Sigma = \Sigma_0 \sigma, \label{3} 
\end{equation}
where $R_0$ and $\Sigma_0$ are the normalization constants of the disk
radius and the surface density, respectively.
Also, we assume that the radial temperature profile has a power-law
distribution with an index, $\beta$,
\begin{equation}
T=T_0 r^{-\beta}. \label{4} 
\end{equation}
In this case, the radial dependences of the sound speed and the
scale height become
\begin{equation}
c_s = c_0 r^{-\beta/2} \ \ {\rm and} \ \  
H = H_0 r^{(3-\beta)/2}, \label{5} 
\end{equation}
where $T_0$, $c_0$, and $H_0=c_0/(GM/R_0^3)^{1/2}$ are the
temperature, the sound speed, and the scale height, respectively, at
$R=R_0$.

Substituting Equations (\ref{3}) and (\ref{5}) into Equation
(\ref{2}), we obtain
\begin{equation}
\frac{j^2}{GMR_0}=r+\left(\frac{H_0}{R_0}\right)^2 r^{(2-\beta)} \left\{\frac{\partial (\mathrm{ln~}\sigma )}{\partial (\mathrm{ln~}r)} -\frac{3+\beta}{2} \right\}.  \label{6}
\end{equation}
Substituting Equation (\ref{6}) into Equation (\ref{1}), we obtain
\begin{eqnarray}
(2-\beta) \frac{\partial (\mathrm{ln~}\sigma )}{\partial (\mathrm{ln~}r)} +\frac{\partial^2 (\mathrm{ln~}\sigma )}{\partial (\mathrm{ln~}r)^2} &+&\left(\frac{R_0}{H_0} \right)^{2}r^{(\beta -1)} \nonumber \\
&-&\frac{1}{2}(3+\beta)(2-\beta) >0, \label{7}
\end{eqnarray}
where we use $r>0$. The disk is rotationally stable when the surface 
density profile satisfies this condition (see also \citealp{Yan10}).

\subsection{Marginally Stable Surface Density Profile for Rotational Instability}
When the disk is marginally stable for rotational instability, the specific angular momentum is constant, \mbox{$\partial j/ \partial R=0$}. 
From Equation (\ref{7}), this condition is equivalent to 
\begin{eqnarray}
(2-\beta) \frac{\partial (\mathrm{ln~}\sigma )}{\partial (\mathrm{ln~}r)} &+&\frac{\partial^2 (\mathrm{ln~}\sigma )}{\partial (\mathrm{ln~}r)^2} +\left(\frac{R_0}{H_0} \right)^{2}r^{(\beta -1)} \nonumber \\
&-&\frac{1}{2}(3+\beta)(2-\beta) = 0 \label{8}.
\end{eqnarray}
The solution of Equation (\ref{8}) is $\sigma_{\mathrm{ms}}$, 
\begin{eqnarray}
\sigma_{\mathrm{ms}} &=& \exp \left[\frac{C_1}{(\beta -2)}r^{(\beta-2)} -\frac{1}{\beta -1}\left(\frac{R_0}{H_0} \right)^{2}r^{(\beta -1)} \right. \nonumber \\
&&\left.  + \frac{3+\beta }{2}\mathrm{ln~}~r +C_2 \right], \label{9} \\
\frac{\partial (\mathrm{ln~}\sigma_{\mathrm{ms}} )}{\partial (\mathrm{ln~}r)}&=& C_1r^{(\beta-2)}-\left(\frac{R_0}{H_0} \right)^{2}r^{(\beta -1)}+\frac{3+\beta }{2}, \label{10} \\
\frac{\partial^2 (\mathrm{ln~}\sigma_{\mathrm{ms}} )}{\partial (\mathrm{ln~}r)^2} &=& (\beta -2) C_1 r^{(\beta-2)} \nonumber \\
&&-(\beta -1) \left(\frac{R_0}{H_0} \right)^{2} r^{(\beta -1)}, \label{11}
\end{eqnarray}
where $C_1$ and $C_2$ are constants of integration. Equation (\ref{9}) represents the marginally stable surface density profile for rotational instability.

\subsection{Rotational Instability in Viscous Disks}
\begin{figure}
\begin{center}
\epsscale{.90}
\plotone{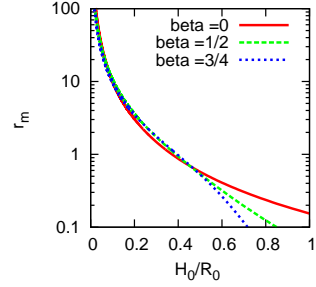}
\caption{Relation between $r_m$ and $H_0/R_0$ obtained by Equation (\ref{21}) for $\beta =0$ (red solid line), $\beta=1/2$ (green dashed line), and $\beta=3/4$ (blue dotted line). }
\label{fig1}
\end{center}
\end{figure}

Here, we examine the critical radius where rotational instability occurs in the similarity solution of the LBP74 model. 
The surface density evolution of viscous disks rotating with the Keplerian velocity is given by  (e.g., \citealp{Pri81}) 
\begin{equation}
\frac{\partial \Sigma}{\partial t} = \frac{3}{R}\frac{\partial}{\partial R} \left[R^{1/2} \frac{\partial}{\partial R} (R^{1/2} \nu \Sigma) \right], \label{12}
\end{equation}
where $t$ is the time and $\nu$ is the coefficient of kinematic viscosity.
In the $\alpha$-viscous disk model \citep{Sha73}, the viscosity $\nu$ is given by  
\begin{equation}
\nu = \alpha c_s H \propto r^{(3/2-\beta)} , \label{13}
\end{equation}
where $\alpha$ is the standard viscous parameter and Equation (\ref{5}) is used.
The similarity solution of Equation (\ref{12}) is
given by \citeauthor*{Lyn74} \citep[1974, ; see also][]{Har98}
\begin{eqnarray}
\Sigma&=& \frac{C}{3\pi \nu_1} T^{-5/(1+2\beta)}r^{(\beta -\frac{3}{2})}\exp{\left[ -r^{(\beta+1/2)} \right]}, \label{14} \\
R_0 &\equiv& T^{2/(1+2\beta)} R_1, \label{15}
\end{eqnarray}
where $R_1$ is an initial radial scale factor, $C$ is a scaling constant, and $\nu_1$ is the coefficient of viscosity at $R_1$.
We note that the disk radius is normalized by $R_0$, which is equal to $R_1$ at $t=0$ and increases with time. While \citet{Har98} used the normalization constant $R_1$, we use $R_0$ instead since it is more convenient for comparison with the marginally stable state discussed in Section 2.2.
The non-dimensional time $T$ is 
\begin{equation}
T \equiv \frac{t}{T_\nu} +1, \label{16}
\end{equation}
where the viscous scaling time, $T_{\nu}$, is 
\begin{equation}
T_\nu \equiv \frac{4}{3(1+2\beta)^2}\frac{R_1^2}{\nu_1}. \label{17}
\end{equation}
In Equation (\ref{15}) we set $R_0$ increasing with time.
We set the surface density normalization constant as 
\begin{equation}
\Sigma_0= \frac{C}{3\pi \nu_1} T^{-5/(1+2\beta)}. \label{18}
\end{equation}
Then the non-dimensional surface density ($\sigma_s$) becomes 
\begin{eqnarray}
\sigma_s &=&r^{(\beta -\frac{3}{2})}\exp{\left[ -r^{(\beta+1/2)} \right]}, \label{19} \\
\frac{\partial (\mathrm{ln~}\sigma_s )}{\partial (\mathrm{ln~}r)}&=& -(\beta+\frac{1}{2})r^{(\beta+1/2)}+(\beta -\frac{3}{2}). \label{20}
\end{eqnarray}
The radius at which the disk becomes rotationally unstable, $r_m$, is given by substituting Equation (\ref{19}) into Equation (\ref{8}) 
\begin{equation}
\left(\frac{R_0}{H_0} \right)^{2}=\frac{5}{2}(\beta+\frac{1}{2})r_m^{3/2}+\frac{1}{2}(6-\beta)(2-\beta)r_m^{(1-\beta)}. \label{21}
\end{equation}
In the outermost part of the disk, $r>r_m$, the similarity profile cannot be maintained because of rotational instability.
Figure 1 shows the relation between $r_m$ and $H_0/R_0$ given by Equation (\ref{21}) with $\beta =0$ (red solid line), $\beta=1/2$ (green dashed line) and $\beta=3/4$ (blue dotted line). 
As discussed in Sect. 4.1 below, $H_0/R_0$ of typical protoplanetary disks ranges from 0.1 to 0.3. 
The critical radius $r_m$ is about 10, 3, and 1 for $H_0/R_0=0.1$, 0.2, and 0.3, respectively (see  Figure \ref{fig1}).

In this paper, we simply assume that the similarity profile connects smoothly to the marginally stable profile, i.e., $\sigma_{\mathrm{ms}}=\sigma_s$ and $\partial \sigma_{\mathrm{ms}}/\partial r=\partial\sigma_s/\partial r$ at $r=r_m$ (e.g., \citealp{Tan07}). Then, from Equations (\ref{9}), (\ref{10}), and (\ref{19})-(\ref{21}), the constants of integration, $C_1$ and $C_2$, are obtained  as
\begin{eqnarray}
C_1 &=& \frac{3}{2}(\beta+\frac{1}{2})r_m^{5/2} +\frac{1}{2}(1-\beta)(6-\beta)r_m^{(2-\beta)},\label{22} \\
C_2 &=&-\frac{C_1}{\beta -2} r_m^{(\beta -2)}+\frac{\beta -6}{2} \mathrm{ln~} r_m +\frac{3(2\beta +3)}{4(\beta -1)}r_m^{(\beta +\frac{1}{2})} \nonumber \\
&&+\frac{(2-\beta )(6- \beta)}{2(\beta -1)}. \label{23}
\end{eqnarray}
This assumption, however, must be examined and we will discuss the more realistic evolution of the surface density profile in Sect.4.3.

%%%%%%%%%%%%%%%%%%%%%%%%%%%%%%%%%%%%%%%%%%%%%%%%%%%%%%%%%%%%%%%
\section{Deviation from the Similarity Profile in the Outer Disks}
In this section, we investigate surface density profiles of the outer region of protoplanetary disks for two different temperature profiles.
In the disks, the temperature profile is controlled by irradiation from the central star and approximately proportional to $r^{-1/2}$, that is, $\beta\sim 1/2$ (e.g., \citealp{Ken87}; \citealp{Chi97}; \citealp{DAl98}).
In the outer disk where the contribution of the irradiation from the central star is weak, or if the disks are irradiated by nearby massive stars, the radial temperature profile becomes roughly isothermal, that is, $\beta\sim 0$ (e.g., \citealp{Rob02}; \citealp{Wal13}).
Here we simply assume a smooth connection from the similarity profile to the marginally stable profile. 
Furthermore, we discuss rotation velocity in the outer region of the disks where the similarity solution is not applicable.

\subsection{Surface Density Profiles}
\subsubsection{Disks Irradiated by the Central Stars ($\beta=1/2$)}
The similarity profile of the surface density of the disks is given by Equation (\ref{19})
\begin{equation}
\sigma_s =r^{-1}\exp{\left( -r \right)}. \label{24} 
\end{equation}
The critical radius where the profile becomes rotationally unstable, $r_m$, is given by Equation (\ref{21})
\begin{equation}
\left(\frac{R_0}{H_0} \right)^{2}=\frac{5}{2}r_m^{3/2}+\frac{33}{8}r_m^{1/2}. \label{25}
\end{equation}
The marginally stable surface density profile for $\beta=1/2$ in Equation (\ref{9}) is 
\begin{equation}
\sigma_{\mathrm{ms}} = \exp{\left[-\frac{2}{3}C_1r^{-3/2} +2\left(\frac{R_0}{H_0} \right)^{2}r^{-1/2}+\frac{7}{4}\mathrm{ln~}~r +C_2 \right]}. \label{26} 
\end{equation}
The smooth connection of the similarity profile to the marginally stable profile at $r=r_m$, gives the constants of integration (Equations (\ref{22}) and (\ref{23})),
\begin{eqnarray}
C_1 &=& \frac{3}{2}r_m^{5/2} +\frac{11}{8}r_m^{3/2}, \label{27} \\
C_2 &=&\frac{2}{3}C_1 r_m^{-3/2}-\frac{11}{4} \mathrm{ln~} r_m -6r_m -\frac{33}{4}. \label{28}
\end{eqnarray}
Figure 2 shows the surface density profile $\sigma$ as a function of radial distance from the central star $r$ in disks with $\beta=1/2$. The black solid line shows the similarity profile. 
Other lines show the marginally stable profiles, which smoothly connect to the similarity profile at $r_m$, for $H_0/R_0 =0.1$ (red solid line), $H_0/R_0 =0.2$ (green long dashed line),  and $H_0/R_0 =0.3$ (blue dotted line). Each square point represents $r_m$. 
It is apparent that for $r>r_m$ the marginally stable profile deviates from
the similarity profile. 
It is also possible to construct marginally stable profiles for $r<r_m$, 
as shown by the dashed lines, but the
similarity profile (black solid line) should be realized in $r<r_m$ in the actual disks. 
For $H_0/R_0=0.1$, the marginally stable profile appears only at the extremely low density region (see the red line). 
This is because the critical radius $r_m$, the location of which is shown by the red square, is much further than the disk radius, $r=1$, of the similarity solution, i.e., $r_m\gg 1$ (see also Figure \ref{fig1}). 
In such disks, deviation from the similarity profile would hardly be detectable. 
However, for thicker disks (for larger $H_0/R_0$), the critical radius $r_m$ decreases, and the
marginally stable profile appears at inner part of the disk where the gas density is higher (see green long dashed and blue dotted lines). 
It is apparent that for $H_0/R_0=0.3$ the marginally stable profile (the blue dotted line) significantly deviates from the similarity profile, keeping relatively high density. 
In such geometrically thick disks, it is expected that the deviation from the similarity profile will be detected by future observations of high sensitivity.

\begin{figure}
\epsscale{.80}
\plotone{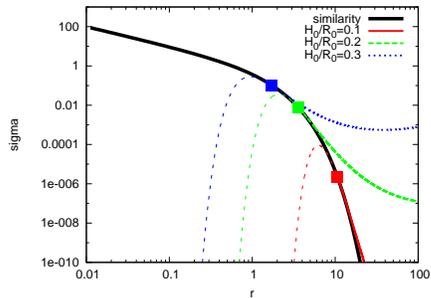}
\caption{Surface density profile $\sigma$ as a function of radial distance from the central star $r$ in disks with $\beta=1/2$. The black solid line shows the similarity profile. 
The other lines show the marginally stable profiles, which smoothly connect to the similarity profile at $r_m$, for $H_0/R_0 =0.1$ (red solid line), $H_0/R_0 =0.2$ (green long dashed line),  and $H_0/R_0 =0.3$ (blue dotted line). Each square point represents $r_m$. }
\label{fig2}
\end{figure}

\subsubsection{Isothermal Disks ($\beta=0$)}
\begin{figure}
\epsscale{.80}
\plotone{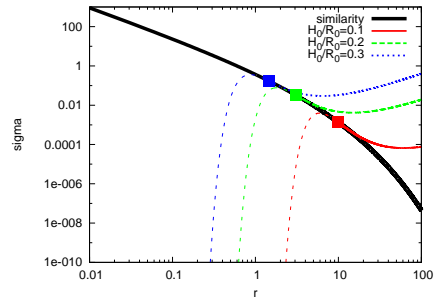}
\caption{Same as Figure \ref{fig2}, but in isothermal disks ($\beta=0$).}
\label{fig3}
\end{figure}

The similarity profile of the surface density of the disks is given by Equation (\ref{19}):
\begin{equation}
\sigma_s =r^{-3/2}\exp{\left( -\sqrt{r} \right)}. \label{29} 
\end{equation}
The critical radius is given by Equation (\ref{21}):
\begin{equation}
\left(\frac{R_0}{H_0} \right)^{2}=\frac{5}{4}r_m^{3/2}+6r_m. \label{30}
\end{equation}
The marginally stable surface density profile of the disks is 
\begin{equation}
\sigma_{\mathrm{ms}} = \exp{\left[-\frac{C_1}{2}r^{-2} +\left(\frac{R_0}{H_0} \right)^{2}r^{-1}+\frac{3}{2}\mathrm{ln~}~r +C_2 \right]}. \label{31} 
\end{equation}
The constants of integration are
\begin{eqnarray}
C_1 &=& \frac{3}{4}r_m^{5/2} +3r_m^{2}, \label{32} \\
C_2 &=&\frac{C_1}{2} r_m^{-2}-3 \mathrm{ln~} r_m -\frac{9}{4}r_m -6. \label{33}
\end{eqnarray}
Figure 3 shows the surface density profile $\sigma$ as a function of radial distance from the central star $r$ in isothermal disks ($\beta=0$). 
As with Figure \ref{fig2}, the black solid line shows the similarity profile and the other lines show the marginally stable profiles, which smoothly connect to the similarity profile at $r_m$, for $H_0/R_0 =0.1$ (red solid line), $H_0/R_0 =0.2$ (green long dashed line),  and $H_0/R_0 =0.3$ (blue dotted line). Each square point represents $r_m$.
Compared to the $\beta=1/2$ case (Figure \ref{fig2}), in isothermal disks the marginally stable profile outside $r_m$ more significantly deviates from the similarity solution. 
Even if $H_0/R_0=0.1$ (red solid line), the marginally stable region appears in high surface density region. 
This is because the similarity profile for $\beta=0$ (black solid line in Figure \ref{fig3}) drops more slowly in the outer region than for $\beta=1/2$ (black solid line in Figure \ref{fig2}). 
In disks with shallower temperature profiles, the deviation from the similarity profile is more easily detectable.
We note that geometrically thin approximation is broken in the region where the surface density increases (see Section 4.2).

\subsection{Rotation Velocity Profiles}
In the marginally stable region ($r\geq r_m$), the specific angular momentum is uniform ($j(r)=j(r_m)$) because $\partial j/ \partial R=0$. 
The rotation velocity ($v_{\phi}$) in $r\geq r_m$ is given by
\begin{equation}
v_{\phi} \equiv \frac{j}{R} = \sqrt{\frac{GM r_m}{R_0}}~ r^{-1}, \label{34}
\end{equation}
which decreases more steeply than the Keplerian profile $v_K \equiv \sqrt{GM/R_0}~ r^{-1/2}$, and thus is slower than the Keplerian velocity. 
In reality, the specific angular momentum $j$ would increases gradually with $r$ (see section 4.3) 
but $(\partial j / \partial R)$ is suppressed.
If $(\partial j/\partial R)$ is significantly reduced, $v_{\phi}$ should be much smaller than $v_K$, and the deviation of rotation velocity from the Keplerian velocity profile can be observable in this region ($r>r_m$).

The deviation of rotation velocity also causes strong drag forces between dust grains and the gas. Dust grains migrate toward the central star and the drag forces make a difference between the surface density profile of the gas and dust grains. Observations indicate the difference (e.g. \citealp{Pan09})  and there are some theoretical works (e.g. \citealp{Tak05}; \citealp{Bir14}) that explain the observations. These theoretical works, however, have not taken into account rotational instability and assume the similarity profiles of the disk gas, that is, they have taken into account the effect of non-Keplerian rotation of the gas for the dust evolution but not for the gas evolution (see Section 4.3 for details). The gas disk profile deviates from the similarity profile in non-Keplerian disks, which affects the dust disk profile. Because future observations of the outer structure of disks, by e.g. ALMA, are expected to unveil both the gas and dust disk structures, theoretical investigations on the gas and dust evolution in non-Keplerian disks is an important topic.

\section{Discussions}
\subsection{Possibility of Observational Detection of Non-similarity Profiles}
In this section, we discuss the possibility of the observational detection of non-similarity profiles in protoplanetary disks, which depends on the value of $H_0/R_0$.
In typical T Tauri disks, the scale height is roughly estimated as
\begin{equation}
\frac{H}{R}\approx 0.1\biggl(\frac{T}{28\mathrm{K}}\biggr)^{1/2}\biggl(\frac{M_*}{M_{\odot}}\biggr)^{-1/2}\biggl(\frac{R}{100\mathrm{AU}}\biggr)^{1/4}, \label{35}
\end{equation}
where we assume $T=280$K at $R=1$AU and adopt $\beta=1/2$,
considering that the temperature profile is controlled by the central
star. In the outer regions where the temperature is as low as that of
the surrounding molecular clouds ($\sim 10-30$K, depending on low- or
high-mass star forming regions),
the temperature profile approaches isothermal ($\beta=0$; see below).
Observationally, $R_0$ ranges from $\sim$15AU to 200AU (\citealp{And09}, \citeyear{And10}), and then $H_0/R_0\sim 0.1-0.18$ from Equation (\ref{35}) if we adopt $M_*=0.5M_{\odot}$.
In this case, $r_m=R_m/R_0$ is roughly 3-10 (see Figure \ref{fig1}). 
For disks with a relatively less massive central star and relatively high temperature, 
$H_0/R_0$ would be close to 0.2. 
For such disks, the deviation of surface density from the similarity profile is so large near $R_m$ that the deviation will be detected by future observations. 
If $H_0/R_0$ is as low as 0.1, it will be difficult to observe the deviation (see Figure \ref{fig2}).
We note that \citet{And07} reported some deviation from the
similarity profiles in their observations of surface density profiles of
protoplanetary disks.
In typical Herbig Ae disks, the temperature is higher ($T\approx 100$K at
$R=10$AU; e.g., \citealp{Dul04}) but the stellar mass is
larger ($M_*\approx 2M_{\odot}$). Therefore, $H_0/R_0$ becomes
smaller and the deviation seems more difficult to observe.

For disks irradiated by nearby massive stars in young
clusters, the external irradiation makes $\beta \simeq 0$ in the outer regions as in the case of protoplanetary disks in the Trapezium Cluster in the Orion Nebula (e.g., \citealp{Rob02}; \citealp{Wal13}).
In this case, the scale height is roughly estimated as
\begin{equation}
\frac{H}{R}\approx 0.2 \biggl(\frac{T}{60\mathrm{K}}\biggr)^{1/2}\biggl(\frac{M_*}{0.5M_{\odot}}\biggr)^{-1/2}\biggl(\frac{R}{100\mathrm{AU}}\biggr)^{1/2}, \label{36}
\end{equation}
where we assume $T=60$K in the outer region and adopt $\beta=0$. 
$H_0/R_0$ approaches about 0.3 if the mass of the
central star is as low as $\sim 0.2M_{\odot}$ (e.g., \citealp{Hil98}). In this case, $r_m=R_m/R_0$ is roughly 1 (see Figure \ref{fig1}), and the deviation from the similarity profile is
possibly observable (see Figure \ref{fig3}). 
In disks near a massive star, however, the gas in the outer region escapes due to photoevaporation, which should be taken into account, (e.g., \citealp{Joh98}; \citealp{Ric00}). 
On the other hand, our result also suggests that the radial pressure gradient force may affect the evolution process due to photoevaporation by surrounding high-mass stars (see Section 4.3).

To summarize, it is expected that deep observations in the near future can detect deviation from the similarity profile of surface density in the outer
regions of the disks whose central stars are less massive and temperature is
relatively high. In the disks heated by irradiation from a nearby
massive star in young star clusters, the deviation will be more easily
observable, but we should take into account the effect of
photoevaporation on the surface density profile in the outer disks.
Observations by facilities with high sensitivity, such as
ALMA, will reveal the surface density profiles of the outer
regions of protoplanetary disks, which tell us the physical
properties of disks with sharp surface density profiles.

\subsection{Break of Geometrically Thin Approximation in The Outer Region}
In this paper, we assume that disks are geometrically thin. 
The assumption, however, breaks in
the outer region of the disks. 
From Equations (\ref{3}) and (\ref{5}), the scale height
is given by
\begin{equation}
\frac{H}{R}=\frac{H_0}{R_0} r^{(1-\beta)/2}. \label{37}
\end{equation}
Figure 4 shows the scale height, $H_m$,  at $R_m\equiv R_0 r_m$ as a
function of $H_0/R_0$, obtained
by Equations (\ref{3}), (\ref{5}), and (\ref{21}) for $\beta =0$ (red solid line), $\beta=1/2$ (green dashed line) and $\beta=3/4$ (blue dotted line). 
Because $H_m/R_m \sim 0.4$ at most, the figure indicates that geometrically thin approximation is appropriate around $R_m$. 
Beyond $r_m$, $H/R$ exceeds unity around the minimum point of the marginally stable surface density profile in Figures \ref{fig2} and \ref{fig3} (for $\beta=1/2$, $r\sim 10^4$, 625, and 100, and for $\beta=0$, $r\sim 100$, 25, and 10 for $H_0/R_0=0.1$, 0.2, and 0.3, respectively) and the geometrically thin approximation is invalid beyond the radius. 
If the gas of the surrounding envelope continues to infall to the disk, the density profile in the outer edge may be more shallower than the self-similar profile. In order to consider this situation, let the density profile be a simple power-law in the whole disk. In this case, $R_m$ would be larger compared to the disk of similarity profiles, resulting in a breakdown of the geometrically thin approximation around $R_m$ (see Equations (\ref{8}) and (\ref{37})). The steep drop of the density at the exponential tail is essential to cause rotational instability.

\begin{figure}
\epsscale{.90}
\plotone{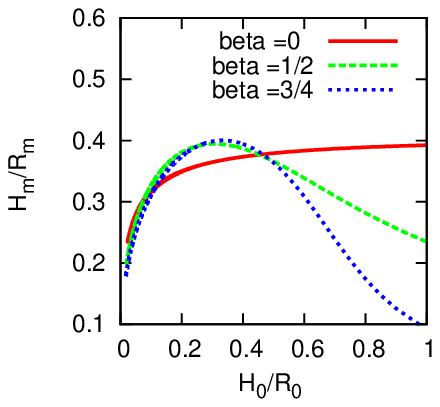}
\caption{Relation between $H_m/R_m$ and $H_0/R_0$ obtained by Equations
 (\ref{3}), (\ref{5}) and (\ref{21}) for $\beta =0$ (red solid line), $\beta=1/2$ (green dashed line) and, $\beta=3/4$ (blue dotted line).  }
\label{fig4}
\end{figure}

\subsection{Evolution of Surface Density Profile in the Outer Region}
In the previous sections, we simply assume that the surface density
profiles connect smoothly from the similarity profile to the marginally stable profile at $r_m$.
In reality, the surface density profile evolves so that
it avoids being rotationally unstable.
In this subsection, we discuss the time evolution of the surface density profile. 
Here, we use only three approximations that disks are axisymmetric and
geometrically thin and that the self-gravitation of disks is negligible.
Now, the equation of continuity is
 \begin{equation}
R\frac{\partial \Sigma}{\partial t}+ \frac{\partial }{\partial R}\left( R\Sigma v_R \right)=0, \label{38}
\end{equation}
the equation of motion in the radial direction is
 \begin{equation}
\frac{\partial v_R}{\partial t} +v_R \frac{\partial v_R}{\partial R} -\frac{j^2}{R^3}= -\frac{1}{\rho}\frac{\partial P}{\partial R}-\frac{GM}{R^2}, \label{39}
\end{equation}
and the equation of angular momentum transfer is
\begin{equation}
2\pi R \Sigma \frac{\partial j}{\partial t} -\dot{M} \frac{\partial j}{\partial R}= \frac{\partial W}{\partial R}, \label{40}
\end{equation}
where $v_R$ is the velocity of the gas in the radial direction, $\dot{M}$ is
the inward mass flux, and $W$ is the viscous torque caused by
turbulent motion.
$\dot{M}$ is defined as 
\begin{equation}
\dot{M} \equiv -2\pi R \Sigma v_R. \label{41}
\end{equation}
$W$ is defined as
\begin{equation}
W \equiv 2\pi R^2 T_{R \varphi} =2\pi R^3 \Sigma \nu \frac{\partial \Omega}{\partial R}, \label{42}
\end{equation}
where $\varphi$ is the azimuthal component in the cylindrical coordinate system and $T_{R\varphi}$ is the $R \varphi$ component of the viscous stress tensor. We assume that the other components of the viscous stress tensor are negligible. 
Using Equations (\ref{41}) and (\ref{42}), Equations
(\ref{38})-(\ref{40}) are 
\begin{eqnarray}
 \frac{\partial \Sigma}{\partial t} &=&\frac{1}{2\pi R} \frac{\partial \dot{M}}{\partial R}, \label{43} \\
\frac{\partial \dot{M}}{\partial t} &=& -2\pi R \Sigma \left(\frac{j^2}{R^3} -\frac{GM}{R^2} -\frac{1}{\rho} \frac{\partial P}{\partial R} \right) \nonumber \\
&&-\frac{\dot{M}^2}{2\pi R^2 \Sigma}\left(1+ \frac{\partial \mathrm{ln}~\Sigma}{\partial \mathrm{ln}~R} -2\frac{\partial \mathrm{ln}~\dot{M}}{\partial \mathrm{ln}~R} \right), \label{44} \\
\frac{\partial j}{\partial t} &=& \frac{1}{2\pi R\Sigma} \left(\dot{M} \frac{\partial j}{\partial R} +\frac{\partial W}{\partial R} \right). \label{45}
\end{eqnarray}
Equation (\ref{43})-(\ref{45}) are differential equations of $\Sigma(R, t)$, $\dot{M}(R, t)$, and $j(R, t)$.

In the previous sections, we ignore the terms that contain second-order of $\dot{M}$ and the terms of $(\partial \dot{M}/ \partial t)$ and $(\partial j / \partial t)$.
In general, we cannot ignore these terms when $\partial j/\partial R$ approaches 0.
Here, in order to qualitatively comprehend how the surface density
profile evolves when $\partial j/\partial R$ approaches 0,
we retain only the term which contains second-order of $\dot{M}$, ignoring the terms $(\partial \dot{M}/ \partial t)$ and $(\partial j / \partial t)$. 
Then Equations (\ref{43})-(\ref{45}) are rewritten as 
\begin{eqnarray}
\frac{\partial \Sigma}{\partial t} &=&\frac{1}{R}\frac{\partial}{\partial R}\left[ \frac{1}{\left( \frac{\partial j}{\partial R} \right)} \frac{\partial }{\partial R}\left\{\nu j \Sigma \left(2- \frac{\partial \mathrm{ln}~ j}{\partial \mathrm{ln}~ R} \right) \right\} \right], \label{46} \\
j^2&=&GMR +\frac{R^3}{\rho}\frac{\partial P}{\partial R} \nonumber \\
&&-\frac{1}{\Sigma^2 \left(\frac{\partial j}{\partial R} \right)^2}\left[\frac{\partial}{\partial R} \left\{\nu j \Sigma \left(2- \frac{\partial \mathrm{ln}~ j}{\partial \mathrm{ln}~ R} \right) \right\} \right]^2 \nonumber \\
&&\left(1+ \frac{\partial \mathrm{ln}~\Sigma}{\partial \mathrm{ln}~R} -2\frac{\partial \mathrm{ln}~\dot{M}}{\partial \mathrm{ln}~R} \right), \label{47} \\
\dot{M} &=& \frac{2\pi}{\left(\frac{\partial j}{\partial R} \right)}\frac{\partial}{\partial R} \left\{ \nu j \Sigma \left(2- \frac{\partial \mathrm{ln}~ j}{\partial \mathrm{ln}~ R} \right) \right\}. \label{48}
\end{eqnarray}
Equation (\ref{46}) is the evolution of the surface density of a geometrically thin viscous disk (similar to Equation (\ref{12}), but without an assumption of the Keplerian rotation) and 
Equation (\ref{47}) is the equation for radial force balance (similar to
Equation (\ref{2})).
Equations (\ref{46}) and (\ref{48}) indicate that when $(\partial j/ \partial R)$ approaches 0 around $R\sim R_m$, the surface density diffuses fast and $\dot{M}$ becomes large,
and then the third term on right side of Equation (\ref{47}) becomes non-negligible. 
However, since the diffusion velocity should be subsonic, the
accretion rate is limited to $|\dot{M}|=2\pi R\Sigma c_s$. 
In this case, $(\partial j/\partial R)$ is roughly estimated from
Equation (\ref{48}) as
\begin{eqnarray}
\frac{\partial j}{\partial R}&=&\frac{2\pi}{\dot{M}}\frac{\partial}{\partial R} \left\{ \nu j \Sigma \left(2- \frac{\partial \mathrm{ln}~ j}{\partial \mathrm{ln}~ R} \right) \right\}  \nonumber \\
&\sim& \frac{2\pi}{2\pi R\Sigma c_s}\frac{1}{R}(2\alpha c_s H j\Sigma) \nonumber \\
&\sim& \alpha c_s\Omega/\Omega_K, \label{49}
\end{eqnarray}
where we adopt $\nu=\alpha c_s H$, $H=c_s/\Omega_K$ and $j=\Omega R^2$. Thus, around $R\sim R_m$, $(\partial j/\partial R)$ is much smaller than that of the disk rotating with
the Keplerian velocity, $\partial j/\partial R\sim v_K$, since $c_s<v_K$ and $\Omega<\Omega_K$.
Therefore, the surface density profile around $R\sim R_m$ will be close to the marginally stable profile, $\sigma_{\mathrm{ms}}$, obtained in Section 3.1. 
However, it is not exactly the same as $\sigma_{\mathrm{ms}}$ since $\partial j/\partial R\neq 0$.
In order to obtain the actual surface density profile, we need to solve Equations (\ref{43})-(\ref{45})  in future work.

\section{Summary}
In this paper, the surface density profile of the outer region of protoplanetary disks is discussed.
We find that the similarity solution is not realized in the outer region due to rotational instability. 
The surface density at the outer region drops with $r$ more slowly than the similarity profile, and
is expected to approach the marginally stable profile.
The deviation from the similarity profile becomes more significant for larger $H_0/R_0$ and for shallower radial temperature profiles.
In order to detect the deviation from the similarity profile, we propose to observe the disks whose central stars are less massive and whose temperature is relatively high. 
In this paper, we calculate the surface density profiles simply assuming the smooth connection of the marginally stable profile to the similarity profile at a critical radius $r_m$ in order to study qualitative behavior of the surface density profile. 
For quantitative comparison to the disk observations, we need to calculate time evolution of the surface density and the rotation velocity simultaneously, as discussed in section 4.3.
Application to other kinds of accretion disks also remains as future work.

%% If you wish to include an acknowledgments section in your paper,
%% separate it off from the body of the text using the \acknowledgments
%% command.

%% Included in this acknowledgments section are examples of the
%% AASTeX hypertext markup commands. Use \url without the optional [HREF]
%% argument when you want to print the url directly in the text. Otherwise,
%% use either \url or \anchor, with the HREF as the first argument and the
%% text to be printed in the second.

\acknowledgments
We thank Hidekazu Tanaka, Takayuki Muto, and Shin Mineshige for their useful comments,
which greatly improved the discussions in the manuscript.
We are also grateful to the referee who helped improve the quality of the manuscript.
This work was partly supported by Grant-in-Aids for Scientific Research, Nos. 23103005 and 25400229.

\end{document}